\documentclass[letterpaper, 10 pt, conference]{ieeeconf}  

\IEEEoverridecommandlockouts                              
\title{\LARGE \bf
Optimism in Games with Non-Probabilistic Uncertainty
}

\author{Jiwoong Lee and Jean Walrand
\thanks{This work is supported by MURI grant BAA 07-036.18.}
\thanks{Jiwoong Lee and Jean Walrand are with the Department of Electrical Engineering and Computer Sciences, University of California at Berkeley, Berkeley, California 94720.
}%
\thanks{\{porce,wlr\}@eecs.berkeley.edu}%
}

\usepackage[cmex10]{amsmath}
\usepackage{graphicx}
\usepackage{psfrag}
\usepackage{array}
\usepackage{threeparttable} 
\usepackage{latexsym}
\usepackage{amsfonts}
\usepackage{amssymb}
\usepackage{amsbsy}
\usepackage{url}
\usepackage[dvips]{epsfig}
\usepackage{color,enumerate,multirow}
\usepackage[normalem]{ulem}
\usepackage[small]{caption}
\usepackage{algorithm2e}
\usepackage{theorem}

\newcommand{\be}{\begin{IEEEeqnarray}{C}}
\newcommand{\ee}{\end{IEEEeqnarray}}
\newcommand{\beno}{\begin{IEEEeqnarray*}{C}}
\newcommand{\eeno}{\end{IEEEeqnarray*}}
\newcommand{\bee}{\begin{IEEEeqnarray}{rCl}}
\newcommand{\eee}{\end{IEEEeqnarray}}
\newcommand{\beeno}{\begin{IEEEeqnarray*}{rCl}}
\newcommand{\eeeno}{\end{IEEEeqnarray*}}

\newtheorem{thm}{Theorem}
\newtheorem{lem}{Lemma}

\newtheorem{define}{Definition}
\newtheorem{example}{Example}

\begin{document}

\maketitle
\thispagestyle{empty}
\pagestyle{empty}

\begin{abstract}

The paper studies one-shot two-player games with non-Bayesian uncertainty.  The players have an {\em attitude} that ranges from optimism to pessimism in the face of uncertainty.  Given the attitudes,
each player forms a belief about the set of possible strategies of the other player. If these beliefs are consistent, one says that they form an {\em uncertainty equilibrium}.  One then considers a two-phase game where the players first choose their attitude and then play the resulting game.  The paper illustrates these notions with a number of games where the approach provides a new insight into the plausible strategies of the players.\end{abstract}


\section{Introduction}
We study a one-shot non-cooperative game of two rational players with non-probabilistic information uncertainty.
Specifically, we assume that the set of possible values of the uncertain parameter is known, but that no prior distribution is available.
Thus, instead of the more traditional Bayesian approach where user maximize their expected reward, here, players have an {\em attitude}
that models their risk-aversion.  An optimistic (respectively, pessimistic) player assumes that the other player will choose a strategy that is beneficial (respectively, detrimental) to her.
A moderately optimistic player makes an intermediate assumption.  However, in contrast with other approaches, we assume that the players choose their
attitude by analyzing the consequences of their choice, instead of assuming that their risk-aversion is pre-determined.

Many researchers have explored non-Bayesian models of uncertainty. Knight \cite{knight1921risk} raised questions about the suitability of probabilistic characterizations of uncertainty in some situations. Allais' parodox and Ellsberg's paradox \cite{ellsberg1961risk} are examples of situations where decision makers violate the expected utility hypothesis. More recently, Binmore \cite{binmore2007making} and Lec and Leroux \cite{le2007bayesian} explored more philosophical questions on inaccuracy, arbitrariness,  and illegitimacy of Bayesianism in games.  The behavioral sociology literature also reports that Bayesian strategies fail to occur in some real world games \cite{varoufakis2001}. A few noteworthy experiments demonstrate a {\em certainty effect} where people prefer less uncertain events, a {\em refection effect} where people respond differently to gain and loss \cite{berg1985preference}, and {\em preference reversals} where people show different valuations when they buy and when they sell the same lottery \cite{chu1990subsidence}.  See also \cite{lim2006model} for a related discussion of the modeling of uncertainty through a family of probability distributions.

Different players may have a different objective in the face of uncertainty. Some popular choices include minimax regret, maximin pessimism or maximax optimism.  Instead of a fixing a player's optimization objective, we allow a rational player to choose somewhere between worst case and best case. We parametrize a player's subjective decision criterion as a convex combination of pessimism and optimism with parameter $\pi$, and we call it a player's \textit{attitude} against uncertainty.  Hurwicz (1951) \cite{hurwicz1951optimality} proposed a similar convex combination criterion for a single agent decision making problem. However, one crucial aspect of this study is that the attitude is not fixed ahead of time. Instead, the players choose their attitude \textit{strategically}. Thus, arbitrariness in choosing a subjective decision criterion disappears while flexibility is maintained. For instance, the players may realize that the only rational attitude is to be optimistic because it is the only Nash equilibrium in a two-stage game where the first stage is to choose the attitude.  More generally, there may be a set of attitudes for each player from which it is not rational to deviate unilaterally. In such a case, the model provides some information about how to behave rationally in the face of uncertainty.

Section \ref{sec:model} develops a model of two non-cooperative players with non-probabilistic parameter uncertainty, and introduces the notions of attitude and uncertainty equilibrium. 
Section \ref{sec:example} presents examples for which the approach provides a new insight into the strategies. 
Section \ref{sec:existence} proves  the existence condition of an uncertainty equilibrium and relates it to a Nash equilibrium of the corresponding full information game.
Section \ref{sec:pessimism} proves that at least one player should not be pessimistic. 
Section \ref{sec:conclusion} concludes the paper.


\section{Uncertainty Equilibrium}
\label{sec:model}

The section defines the model of game with uncertainty. It then introduces the notion of uncertainty equilibrium for players that have specific attitudes. The section then defines the two-phase game. First, we define a reference game with full information.

\begin{define}[Certainty Game $\mathcal{G}_{o}$]
\ \\
Two non-cooperative, selfish and rational players $i=1,2$ and $j = 3-i$ play a game with strategies $x:=(x_{1}, x_{2}) \in X_{1,o} \times X_{2,o}$, where $X_{i,o} \subset \mathbb{R}$ is $i$'s closed bounded strategy interval. Player $i$ has type $\theta_{i} \in \mathbb{R}$. 
The reward of player $i$ is real-valued $u_i(x, \theta_i)$.
This is a full information game with common knowledge about 
$u_i$, $X_{i,o}$, and $\theta_i$  for all $i$.
We assume that this game is such that $u_i(x, \theta_i)$ is continuous in $(x, \theta_{i})$, has a unique maximizer
$x_i(x_j, \theta_i)$ for every $(x_j, \theta_i)$, and has at least one pure Nash equilibrium.  
\end{define}

We now consider the game with uncertainty about the opponent's type. 
\begin{define}[Uncertainty Game $\mathcal{G}$]
\ \\
Player $i$ knows her own true type $\theta_{i}$ but only that $\theta_{j} \in \Theta_{j}$ for $j=3-i$, where $\Theta_{j}$ is a closed bounded real interval, and this is common knowledge.  To avoid triviality, $\Theta_j$ is assumed to be of non-zero length unless specified otherwise. 
\end{define}

The goal of the paper is to study the notion of equilibrium in such a situation.  Our approach is non-Bayesian. That is, we do assume neither a known posterior distribution of the parameters nor the existence of a common prior distribution.

We start with a simple approach to refine the set of rational strategies. Assume that it is known that player $i$ chooses $x_i \in X_i$. It may be reasonable to believe that player $j$ will choose a strategy $x_j(x_i, \theta_j)$ for some $x_i \in X_i$. Since player $i$ does not know $\theta_j$, she may then believe that player $j$
chooses $x_j \in \phi_j(X_i)$ where
\be
\phi_j(X_i) := \{ x_j (x_i, \theta_j) \mid x_i \in X_i, \theta_j \in \Theta_j\}.
\ee

These considerations lead to the following definition.

\begin{define} \label{d.consistent}
The sets $X_1, X_2$ are {\em consistent} if $X_j = \phi_j (X_i)$ for $i = 1, 2$ and $j = 3 - i$.
\end{define}

The consistent sets form a product space of strategies beyond which no rational player plays. Although the sets $X_i$ are smaller than the original strategy spaces $X_{i,o}$, they may be large and provide little recommendation on the strategies the players should choose.  Moreover, one may question whether the players will choose strategies in the consistent sets.

\subsection{Optimism and Pessimism}

We now develop a different formulation of the game that considers the
{\em attitudes} $\pi = (\pi_{1}, \pi_{2}) \in [0, 1]^2$  of players in the face of uncertainty. 
\begin{define}[Game with Attitudes $\pi$: $\mathcal{G(\pi)}$]
\ \\
If it is known that player $j$ chooses $x_j \in X_j$, then player $i$ chooses $x_i \in X_{i, o}$ to maximize
\[
f_i(x_i, X_{j}, \theta_i, 1) := \max_{x_j \in X_j} u_i (x, \theta_i)
\]
if she is optimistic and to maximize
\[
f_i(x_i, X_{j}, \theta_i, 0) := \min_{x_j \in X_j} u_i (x, \theta_i)
\]
if she is pessimistic.
In general, for $0 \le \pi_{i} \le 1$, if player $i$ has attitude $\pi_i$, she chooses $x_{i}  \in X_{i, o}$ to maximize 
\begin{eqnarray}
&&f_i(x_i, X_j, \theta_i, \pi_i) \nonumber \\
&&~:= \pi_i \max_{x_j \in X_j} u_i(x, \theta_i) + (1-\pi_i) \min_{x_j \in X_j} u_i(x, \theta_i).
\end{eqnarray}
\end{define}

We primarily study a discrete attitude space $\pi_{i} \in \{0, 1\}$, and later use the continuous attitude space $\pi_{i} \in [0,1]$ in developing the notion of robust attitude.

Designate by $r_{i} (X_j, \theta_i; \pi_{i})$
the set of maximizers of $f_{i}(x_i, X_{j}, \theta_i , \pi_{i})$. 
\be
	r_i(X_j, \theta_i, \pi_i) := \arg \max_{x_i \in X_{i,o}} f_i(x_i, X_j, \theta_i, \pi_i).
\ee
Since player $j$ does not know $\theta_i$, she assumes that $x_i \in \psi_i (X_j; \pi_{i})$ where
\be
	\psi_i(X_j; \pi_{i}) := \bigcup_{\theta_i \in \Theta_i} r_i (X_j, \theta_i; \pi_{i}).
\ee

\subsection{Uncertainty Equilibrium}

We then have the following definition.
\begin{define}[Uncertainty Equilibrium of $\mathcal{G}(\pi)$]
\ \\
The pair of sets $(X_1, X_2)$ is an {\em uncertainty equilibrium for players with attitudes $\pi$},  if $X_i = \psi_i (X_j; \pi_{i})$ for $i = 1, 2$ and $j = 3 - i$.  
\end{define}

Moreover, if the uncertainty equilibrium is unique, we consider that player $i$ plays $x_i \in r_{i}(X_{j}, \theta_{i}; \pi_{i})$ to maximize her {\em interim anticipated reward} $f_i (x_i, X_{j}, \theta_i, \pi_i)$. 
If the corresponding $x_i$ is unique and equal to $x_i (\theta_i, \pi)$, it results in {\em actual (ex-post) rewards}
$U_i  := u_i (x_i (\theta_i, \pi), x_j (\theta_j, \pi), \theta_i)$. If the context is clear, we simplify as  $U_i(\pi)  := u_i (x_i (\pi), x_j (\pi), \theta_i)$ where $x_{i}(\pi) = x_{i}(\theta_{i},\pi)$.

\subsection{Attitude Game}

Is it preferable to be optimistic or pessimistic?  To answer this question, we consider a two-stage game. 

\begin{define}[Attitude Game $\mathcal{A}$]
\ \\
In the first stage, the players choose their attitudes $(\pi_{1}, \pi_{2}) \in \{0, 1\}^2$. In the second stage, they play $\mathcal{G}(\pi)$ and get
the rewards $U_i (\pi)$.  
\end{define}

If $\pi=(0, 0)$ is a unique Nash equilibrium for the two-stage game, we conclude that the players should be pessimistic 
Moreover, the analysis then specifies precisely how they should choose their second stage strategy.  The situation is similar if any $\pi \in \{0, 1\}^2$ is a unique Nash equilibrium attitude. 
{A player $i$'s attitude $\pi_{i}^{*}$ is said to be {\em dominant} if for any $\pi_{j}$ and $\theta_{j}$, $j=3-i$,

\[
	U_{i}(\pi_{i}^{*}, \pi_{j}) \ge U_{i}(\pi_{i}, \pi_{j})
\]
for all $\pi_{i}$.} 

In contrast with traditional approaches, we do not consider that players have a fixed attitude (as a  {\em type}).
Instead, they decide whether to be optimistic or not given the game.  They choose their attitudes by
analyzing the game instead of being driven by a preordained risk aversion.

As we show in the following sections, there are games where this approach enables to rationalize specific strategies under uncertainty.

\section{Examples}
\label{sec:example}

The first example is a game with negative externality. In this game, the players should be optimistic even when they are uncertain about the opponent's type. The second example is a Cournot duopoly game \cite{cournot1838recherches}  with uncertainty. For this game, we study conditions for the existence of dominant attitudes, and robust attitudes.
For clarity, the algebraic derivations are in the appendix.

\subsection{A Game with Negative Externality}
{
Consider two players $i=1,2$ consume resource $x_{i} \in [0,1]$ to gain benefit but also the consumption degrades the quality of the environment which affects both players.  The player's reward is defined to be the benefit minus the degradation of the environment quality.
The benefit is assumed to be proportional to the consumption. The environment degrades exponentially in sum of players' consumption ($\exp \{ x_{1} + x_{2} \})$, via scaling factor $\exp \{ -\theta_{i} \}$, where ${\theta_{i}^{{-1}}}$ captures $i$'s susceptibility to the environmental degradation. $\theta_{i}$ is private information. 
}
$x_i \in [0, 1], \theta_i \in [\alpha, \beta]$ for some $0 < \alpha < 2 \alpha <  \beta < 1$. 
Operator $i$'s reward is 
\[
u_i(x, \theta_i) = x_i - \exp \{ - \theta_i + x_i + x_j\} .
\]
(One may add a constant to make the rewards positive.)

\begin{thm}
\label{thm:interferencegame}
Players should be optimistic and choose the consumption levels $x_{i} = \theta_{i} - \alpha/2$ for $i = 1, 2$.
In contrast, if $\theta_{1}, \theta_{2}$ are fully known and $\theta_{1} < \theta_{2}$, then the only Nash strategy is $(x_{1}, x_{2}) = (0, \theta_{2})$. 
\end{thm}

For this game, the only consistent sets (see Definition \ref{d.consistent}) are $X_1 = X_2 = [0, \beta]$, which provides little information about the strategies of the players.

\subsection{Cournot Duopoly Game}
\label{sec:cournot}

\subsubsection{Full Information Case}

For $i = 1, 2$, selfish and rational player $i$ produces a non-negative quantity $x_i$ of homogeneous items with a non-negative production cost $\theta_i \in [0, 1/2]$ per item. The selling price per item is $(1 - x_1 - x_2)^+$  where $y^+ = \max\{y, 0\}$ for $y \in \Re$.  Accordingly, the reward (profit) of player $i$ is $u_i (x, \theta_i)$ defined as follows:
\be \label{e.g1}
    u_i(x,\theta_i) := x_i (1 - x_1 - x_2)^+ - \theta_i x_i
\ee
where $x = (x_1, x_2)$.

Player $i$'s strategy is the quantity $x_i$ to produce. The value of $x_i$ that maximizes $u_i(x, \theta_i)$ is $x_i = (1-\theta_i-x_j)/2$, for $ i = 1,2$ and $j = 3 - i$.
The unique solution of these equations is the Nash equilibrium  $x^\star:=(x_1^\star, x_2^\star)$ where
\be \label{eq:cournot_ne}
x_i^\star = (1 - 2\theta_i + \theta_j)/3.
\ee
The corresponding utilities are 
\be \label{eq:cournot_ne_util} u_{i}^\star = {x_i^\star}^2.\ee
Note that the pair $x = (x_1, x_2)$ that maximizes $u_{social} := \sum_{i=1,2} u_i (x, \theta_i)$ is $( (1 - \theta_1)/2, 0)$ when $\theta_1 < \theta_2$.  This ``social optimum'' is quite different from the Nash equilibrium. There 
\be \label{eq:cournot_social_util} u_{social} = (1-\theta_1)^2/4.\ee
\subsubsection{Bayesian Uncertainty Case}
In a Bayesian model, one assumes that $\theta_1$ and $\theta_2$ are independent with known distributions; each player $i = 1, 2$ knows $\theta_i$ and only the distribution of $\theta_j$ for $j = 3 - i$, and this is common knowledge.
In that case,
\[
E[u_1 (x,\theta_1) | x_1, \theta_1] = x_1 (1 - x_1 - E[x_2|x_1, \theta_1] ) - \theta_1 x_1
\]
and this expression is maximized by 
\beno
\label{eqn:bayes}
x_1 =  (1 - E[x_2|x_1, \theta_1] - \theta_1)/2 = (1 - E(x_2) - \theta_1)/2.
\eeno
The last expression follows from the observation that $x_2$ is only a function of $\theta_2$ which is independent of $\theta_1$.
Consequently, for $i = 1, 2$,
\beno
\label{eqn:bayes1}
E(x_i) =  (1 - E(x_j) - \mu_i)/2 \mbox{ where } \mu_i := E(\theta_i).
\eeno
Solving this system of two equations, we find
\[
E(x_1) = (1 - 2\mu_1 + \mu_2)/3 \mbox{ and } E(x_2) = (1 - 2\mu_2 + \mu_1)/3.
\]
Accordingly, for $i = 1,2$,
\be
\label{eqn:bayesian}
x_i =  (2-3\theta_i-\mu_i+2\mu_j)/6 \mbox{ where } j = 3 - i.
\ee
This solution is a unique Bayesian Nash equilibrium. Note that player $i$'s strategy maximizes her \textit{interim expected utility} $E[u_i(x,\theta_i) | x_i, \theta_i]$, rather than the \textit{ex post utility} $u_i(x,\theta_i)$, which $i$ cannot compute. 

\vspace{0.1in}
\subsubsection{Game with Attitudes}

One assumes that, for $i = 1, 2$, player $i$ knows $\theta_i$ but only that $\theta_j \in \Theta_j := [\alpha_j, \beta_j]$ for $j = 3 - i$
where $\beta_j \leq 1/2$.
This is common knowledge. Moreover, player $i$ has attitude $\pi_i \in [0, 1]$.  The following result is shown in the appendix.

\begin{thm}
\label{thm:cournot}
The unique uncertainty equilibrium with attitudes $\pi$  is the pair of intervals $\mathcal{B}[s_i, t_i] := [s_i - t_i/2, s_i + t_i/2]$ for $i=1,2$, where
\be
\label{eqn:cournot_ue}
	s_i = \frac{1}{3}\Delta_j \pi_i - \frac{1}{6}\Delta_i \pi_j + \frac{1}{12}(4-3\beta_i-5\alpha_i+4\alpha_j)
\ee
and $\Delta_i:=\beta_i-\alpha_i$ and $t_i = (\beta_i-\alpha_i)/4$. The strategies that maximize the interim anticipated rewards are 
\be
\label{eqn:cournot_be}
	x_i^*(\pi) = \frac{1}{3}\Delta_j\pi_i - \frac{1}{6}\Delta_i\pi_j + \lambda_i,
\ee
where $\lambda_i = (2-\alpha_i+2\alpha_j-3\theta_i)/6$. 
\end{thm}

Based on this analysis, one considers the two-stage game $\mathcal{A}$.
The following lemma is proved in the appendix.

\begin{lem}[Dominant attitude]
\label{lem:dominance}
\ \\
Let ${\underline \theta}_i:= \frac{1}{3}(2-\beta_i+4\alpha_j-2\beta_j)$ and ${\overline \theta}_i:= \frac{1}{3}(2-\alpha_i+4\beta_j-2\alpha_j)$. Assume that the attitude space is discrete $\Pi = \{0,1\}$.
\begin{enumerate}
\item	If $\theta_i \le {\underline \theta}_i$, then optimism is a dominant strategy for player $i$.
\item  If $\theta_i \ge {\overline \theta}_i$, then pessimism is a dominant strategy for player $i$. 

\item If $\underline{\theta_i} < \theta_i < \overline{\theta}_i$, then there is no dominant strategy for player $i$.

\end{enumerate}
\end{lem}

In particular, if $\beta_{i} < 1/3$ for $i =1, 2$ (i.e., if the unit production costs are sufficiently low), both players should be optimistic. 

The game is said to be {\em symmetric} if $u_{1} = u_{2}$ and $\Theta_{1} = \Theta_{2}$.  
The following result corresponds to a symmetric game.
\begin{thm}
\label{thm:pd}
Consider the game $\mathcal{A}$ with $\Theta_1 = \Theta_2 =[\alpha,\beta]$ where $\beta > \alpha$. 
\begin{enumerate}
	\item	$(PP)$ is never a Nash equilibrium. 
	\item	$(PP)$ is pareto efficient. 
	\item $(PP)$ is pareto superior to $(OO)$.
	\item $O$ is the dominant strategy if $\beta \le \max(1/3, 2\alpha)$. Then $(OO)$ is the only Nash equilibrium.
\end{enumerate}
Together with 1), 2), and 3), the condition in the last part makes the attitude game a Prisoner's Dilemma. 
The last condition requires that the costs are not too large.
\end{thm}

\subsubsection{Robust attitude}
As we observed from the previous example, game $\mathcal{A}$ may not have a dominant attitude for player $i$. In such a case, player $i$ may prefer a strategy that guarantees the largest minimum ex-post reward.  That is, player $i$  might seek the {\em robust attitude} $\pi_{i}^{\sharp} \in [0,1]$ defined by
\[
	\pi_{i}^{\sharp} := \arg \max_{\pi_{i}} \min_{\pi_{j}} u_{i}(x_{i}(\theta_{i}, \pi), x_{j}(\theta_{j}, \pi), \theta_{i}).
\]

\begin{thm}
\label{thm:cournot_robust}
The robust attitude of Cournot duopoly does not coincide with pessimism and is given by 
\[
	\pi_i^\sharp = \min(1,{(2-3\theta_i - \beta_i + 2\alpha_j)}/{4\Delta_j})
\]
for $\Delta_j > 0$. Consequently, $\pi_i^\sharp > 0$, except for a singular case  $\alpha_j=0$ {\it and} $\theta_i=\beta_i=1/2$.
\end{thm}

\begin{example} Let $\beta:=\max(\beta_i,\beta_j)$. Then if $\beta \le 1/4$, $\pi_i^\sharp = \pi_j^\sharp = 1$. That is, when costs are sufficiently small, the robust strategy is optimism. To see this,
note that $
\pi_i^\sharp = \min(1, (2-3\theta_i  -\beta_i + 2\alpha_j)/4(\beta_j-\alpha_j)) \ge \min(1, (2-4\beta)/4\beta)) = 1.
$ 
\end{example}

%
%
%
%
%

\section{Existence of Uncertainty Equilibrium and its Relation to Nash Equilibrium}
\label{sec:existence}

This section provides a condition for the existence of an uncertainty equilibrium. 

\begin{thm}[Existence of Uncertainty Equilibrium]
\label{thm:existence}
\ \\
{
Assume $r_{i}(X_{j}, \theta_{i}, \pi_{i})$ is single-valued and continuous in $X_{j}, \theta_{i}$ and $\pi_{i}$. 
}
Then there exists an uncertainty equilibrium $(X_1^*(\pi), X_2^*(\pi))$.
\end{thm}

At an uncertainty equilibrium $(X_1^*(\pi), X_2^*(\pi))$, $i$'s best response is
\[
x_i^*(\pi) = r_i(X_j^*(\pi), \theta_i, \pi_i).
\]
From the proof of Theorem \ref{thm:existence}, note there is one-to-one correspondence between $x_i^*(\pi)$'s and $X_i^*(\pi)$'s via $r_i$'s. In particular, if $\Theta_i$ is a singleton, then $X_i^*(\pi) = x_i^*(\pi)$. This observation is stated in the next theorem.

\begin{thm}
\label{thm:coincidence} 
Under the assumptions of Theorem  \ref{thm:existence}, $\mathcal{G}(\pi)$'s uncertainty equilibrium $(X_1^*(\pi), X_2^*(\pi))$ coincides with game $\mathcal{G}_{o}$'s Nash equilibrium $(x_1^\star, x_2^\star)$ if $\Theta_i= \{\theta_i\}$ for $i = 1, 2$, irrespective of  $\pi$. 
\end{thm}

\section{At least one player does not prefer pessimism}
\label{sec:pessimism}

We identify conditions when pessimism cannot be dominant for both players.

The first theorem proves this for the non-symmetric Cournot duopoly game. The following theorem is for a more general utility structure of symmetric games.

\begin{thm}
\label{thm:no_pessimism}
Both Cournot duopoly players cannot simultaneously have pessimism as their dominant attitude.
\end{thm}

Now we consider a more general utility function case.

\begin{thm}
\label{thm:no_pessimism}
Consider a symmetric game where $u_{i}$ is strictly monotonic in $x_{j}$ and $r_{i}(x_{j}, \theta_{i})$ is single valued and
strictly monotonic in $x_{j}$ and $\theta_{i}$.  Then pessimism cannot be a dominant attitude for any of the two players. 
\end{thm}

\section{Conclusions}
\label{sec:conclusion}
This paper proposes a framework to analyze two-player games with non-probabilisitc information uncertainty. The formulation allows a rational player to choose an attitude against uncertainty
characterized by a degree of optimism. Corresponding to a pair of attitudes, we define an uncertainty equilibrium as a pair of sets of strategies from which rational players would not depart unilaterally. Under some assumptions, this concept coincides with the traditional Nash equilibrium when there is no uncertainty. We then define a two-phase game where players first choose their attitude. Finally, we illustrate the framework with an investment game and a Cournot duopoly game with uncertainty. We show that the framework may identify uniquely the strategies of the players.

\section*{Appendix}

\subsection{Proof of Theorem \ref{thm:interferencegame}}
The partial derivative with respect to $x_i$ is $1 - \exp \{- \theta_i + x_i + x_j\}$, which is positive for
$x_i < \theta_i - x_j$ and negative for $x_i > \theta_i - x_j$.
Accordingly, the best response $x_i(x_j)$ is
$x_i(x_j) = [\theta_i - x_j]^+$. 
If $\theta_i < \theta_j$, the only Nash equilibrium is then 
$x_i = 0, x_j = \theta_j$.  The outcome of the game is very sensitive to the order of the parameters. 

Assume $i$ knows that $x_j \in X_j$.  For $z \in \mathbb{R}$, define $[z]_0^1 := \min \{\max\{z, 0\}, 1\}$. Then, if $i$ is optimistic, she maximizes $x_i - \exp \{- \theta_i +  x_i + \alpha_j\}$ where $\alpha_j = \min X_j$.  Thus,
\[
x_i = [\theta_i - \alpha_j]_0^1 \in [[\alpha - \alpha_j]_0^1, [\beta - \alpha_j]_0^1].
\]
Also, if $i$ is pessimistic, she maximizes
$x_i - \exp \{- \theta_i + x_i + \beta_j\}$
where $\beta_j = \max X_j$. Thus,
\[
x_i = [ \theta_i - \beta_j]_0^1 \in [[\alpha - \beta_j]_0^1, [\beta - \beta_j]_0^1].
\]

Suppose both players are optimistic.  Then the
only uncertainty equilibrium is $X_i = X_j = [a, b]$ where
$
a = \alpha - a \mbox{ and } b = \beta - \alpha.
$
Hence 
$X_i = X_j = [\frac{\alpha}{2}, \beta - \frac{\alpha}{2}]$.
Consequently,
$
x_i = \theta_i - \frac{\alpha}{2}
$
and
\[
U_i(1, 1) := \theta_i - \frac{\alpha}{2} - \exp\{\theta_j - \alpha\}.
\]

Second, suppose both players are pessimistic.  Then
the only consistent sets are $X_i = X_j = [a, b]$ where
$
a = \alpha - b \mbox{ and } b = \beta - b.
$
Hence,
$
X_i = X_j = [\alpha - \frac{\beta}{2}, \frac{\beta}{2}].
$
Consequently,
$
x_i = \theta_i - \frac{\beta}{2}
$
and
\[
U_i (0, 0) := \theta_i - \frac{\beta}{2} - \exp \{\theta_j - \beta\}.
\]

Third, suppose that player 1 is optimistic and player 2 is pessimistic. In that case, the only consistent sets are $X_1 = [a_1, b_1]$ and
$X_2 = [a_2, b_2]$ where
$
a_1 = [\alpha - a_2]_0^1, b_1 = [\beta - a_2]_0^1, a_2 = [\alpha - b_1]_0^1, b_2 = [\beta - b_1]_0^1.
$
Hence,
$
X_1 = [\alpha, \beta] \mbox{ and } X_2 = \{0\}.
$
Consequently,
$
x_1 = \theta_1 \mbox{ and } x_2 = \theta_2 - \beta,
$
so that
\[
U_1(1, 0) := \theta_1 - \exp \{\theta_2 - \beta\}.
\]
By symmetry,
\[
U_1(0, 1) := \theta_1 - \beta - \exp \{ \theta_2 - \beta \}.
\]
By inspection, we see 
\[
U_1(1, 0) \geq U_1(0, 0) \mbox{ and } U_1(1, 1) > U_1(0, 1).
\]
Thus, optimism is a dominant strategy for player 1.  By symmetry, it is 
also dominant for player 2.

\subsection{Proof of Theorem \ref{thm:cournot}}
The proof goes in following steps: First we define the uncertainty set as a ball. Then we show the ball's radius is constant. Finally we show the center of the ball is fixed at equilibrium. 
Note $u_i$ is negatively affine in $x_j$. Let $X_{o}=[0,1/2]$ be the strategy space. Thus
$\inf X_j = \arg \sup_{x_j \in X_j} u_i(x, \theta_i)$ and $\sup X_j = \arg \inf_{x_j \in X_j} u_i(x, \theta_i)$. Define
\[
	h_i(X_j, \pi_i) = \pi_i \inf X_j + (1-\pi_i) \sup X_j.
\]
Then $f_i(x_i, X_j, \theta_i, \pi_i) 
	=u_i(x_i, h_i(X_j, \pi_i), \theta_i)$.
From the first order condition and definition, $i$'s best response to $X_j$ becomes
\beno
	r_i(X_j, \theta_i, \pi_i) = (1-h_i(X_j, \pi_i)-\theta_i)/2.
\eeno
This yields 
\beeno
\sup X_i &=&  (1-r_i(X_j, \pi_i)-\alpha_i)/2\\
\inf X_i &=&  (1-r_i(X_j, \pi_i)-\beta_i)/2.
\eeeno

Now let $X_i^*=\mathcal{B}[s_i, t_i]$ for $i=1,2$ and $j\neq i$ where $\mathcal{B}[s,t]$ is a closed ball or radius $t$ centered at $s$. Then
\beno
t_i = (\sup X_i^* - \inf X_i^*)/2 = \Delta_i/4,
\eeno
where $\Delta_i:=\beta_i-\alpha_i$. This is independent of $X_i^*,X_j^*,\theta_i, \theta_j$. 
Now since $\sup X_j^* = s_j + t_j$ and $\inf X_j^*=s_j - t_j$,
\beeno
	h_i(X_j^*, \pi_i) &=&  s_j + t_j(1-2\pi_i).
\eeeno
Define $\sigma_i := (\alpha_i + \beta_i)/4$,. Then 
\beeno 
s_i &=& (\sup X_i^* + \inf X_i^*) / 2 \\
 &=& (1-r_i(X_j^*)) / 2 - \sigma_i = (1-s_j-t_j(1-2\pi_i))/2 - \sigma_i
\eeeno
for $i=1,2$. We have two equations relating $s_i$ and $s_j$. By solving algebra, we get \eqref{eqn:cournot_ue}.
$i$'s best response at uncertainty equilibrium $x_i = r_i(X_j^*, \theta_i, \pi_i)$ becomes \eqref{eqn:cournot_be}.
$\mathcal{B}[s_i, t_i]$ then is uniquely determined by given $(\pi, \theta_i, \Theta_i, \Theta_j)$. To show its existence, it is sufficient to show $\mathcal{B}[s_i, t_i] \subset {X_o}$. To see this, it is straightforward to verify $\min s_i + t_i \le \sup X_o$ and $\max s_i - t_i \ge \inf X_o$ for all combinations of $\pi, \Theta_1, \Theta_2$.

\subsection{Proof of Lemma \ref{lem:dominance}}
\begin{enumerate}
\item We need to find a condition, without loss of generality, such that (i) $u_1(OO) \ge u_1(PO)$ and (ii) $u_1(OP) \ge u_1(PP)$ for every $\theta_2 \in [\alpha_2, \beta_2]$. By algebra,
\begin{IEEEeqnarray*}{rCl}
	u_1(OO) - u_1(PO)
	\ge \Delta_2 [{\underline \theta}_1 - \theta_1]/12,
\end{IEEEeqnarray*}
which is non-negative for all $\theta_1 \le {\underline \theta}_1:= \frac{1}{3}(2-\beta_1+4\alpha_2-2\beta_2)$ and for all $\theta_2$. 
(ii) is immediate because 
\begin{IEEEeqnarray*}{rCl}
	u_1(OP) - u_1(PP) 
	\ge u_1(OO) - u_1(PO).
\end{IEEEeqnarray*}

\item Similar development yields $\theta_i \ge {\overline \theta}_i$ where ${\overline \theta}_i := \frac{1}{3}(2-\alpha_i-2\alpha_j+4\alpha_i)$.
\end{enumerate}

\subsection{Proof of Theorem \ref{thm:pd}}
\begin{enumerate}
	\item	We show at least one player always have an incentive to deviate from $(PP)$. This part of the theorem is true even for non-symmetric $\Theta_1$ and $\Theta_2$. Define $u_i(\pi) := u_i(x_i^*(\pi), x_j^*(\pi), \theta_i)$ and $\Delta = \beta-\alpha$. Suppose player 1 does not have the incentive to deviate from (PP).  That is, $u_1(PP) \ge u_1(OP)$. Then we prove by showing $u_2(OP) > u_2(PP)$.  From the proof of Lemma \ref{lem:dominance}, $u_1(PP) \ge u_1(OP)$ is equivalent to $3\theta_1 \ge 2- 3 \alpha - 2\beta + 6\theta_2$. Then, $u_2(PO)-u_2(PP) = \frac{\Delta_1}{36}(2-3\alpha-2\beta+6\theta_1-3\theta_2) \ge \frac{\Delta_1}{36}(6-9\alpha-6\beta+ 9\theta_2) > 0$.  The last inequality comes from the boundary condition $0\le \alpha \le \theta_i \le \beta \le 1/2$.
	\item We show that a rival player's optimistic attitude is always detrimental: $36(u_1(PP) - u_1(PO)) = \Delta(6x_1(PP)) + \Delta(6-6\theta_1-6x_1(PP)-6x_2(PP) - \Delta) > 0$. We can similarly show $36(u_1(OP)-u_1(OO))> 0$. At $(PP)$, suppose one player has incentive to change to $O$. That change hurts the ex post utility of the other player. This concludes $(PP)$ is pareto efficient. 
	\item We need to show $u_i(PP) > u_i(OO)$. To see this, $36(u_i(PP)-u_i(OO))=12\Delta x_1(PP) - \Delta(6-6\theta_1 -6x_1(PP) - 6x_2(PP)-2\Delta) = \Delta(2+2\alpha+2\beta-3\theta_1-3\theta_2) \ge 0$. 
	\item	If $\beta \le \max(1/3, 2\alpha)$, then $\underline \theta_i \ge \beta \ge \theta_i$ for all $i$, and importantly, this fact becomes a common knowledge. From Lemma \ref{lem:dominance}, $O$ is the dominant strategy. Together with 1), 2) and 3), this constitutes a Prisoner's Dilemma game.
\end{enumerate}

\subsection{Proof of Theorem \ref{thm:cournot_robust}}
$u_i(q^*(\pi), \theta_i)$ is non-increasing in $\pi_j$ for all possible combinations of parameters. Thus $u_i$ is minimized at $\pi_j = 1$. $u_{i}$ is convex in $\pi_{i}$. From the first order condition, the result is immediately obtained.

\subsection{Proof of Theorem \ref{thm:existence}}
Since $r_{i}$ is continuous in $\theta_{i}$, and $\Theta_{i}$ is a bounded and closed interval, $X_i$ is a closed interval. Let $X_i = [\underline{x}_i, \overline{x}_i] \subset X_{i,o}$, $\underline{x}_i \le \overline{x}_i$. We define a map $\phi(\underline{x}_i, \overline{x}_i) = (\underline{x}'_i, \overline{x}'_i)$ such that
\begin{figure}[h]
	\centering
	\includegraphics[width=2.3in]{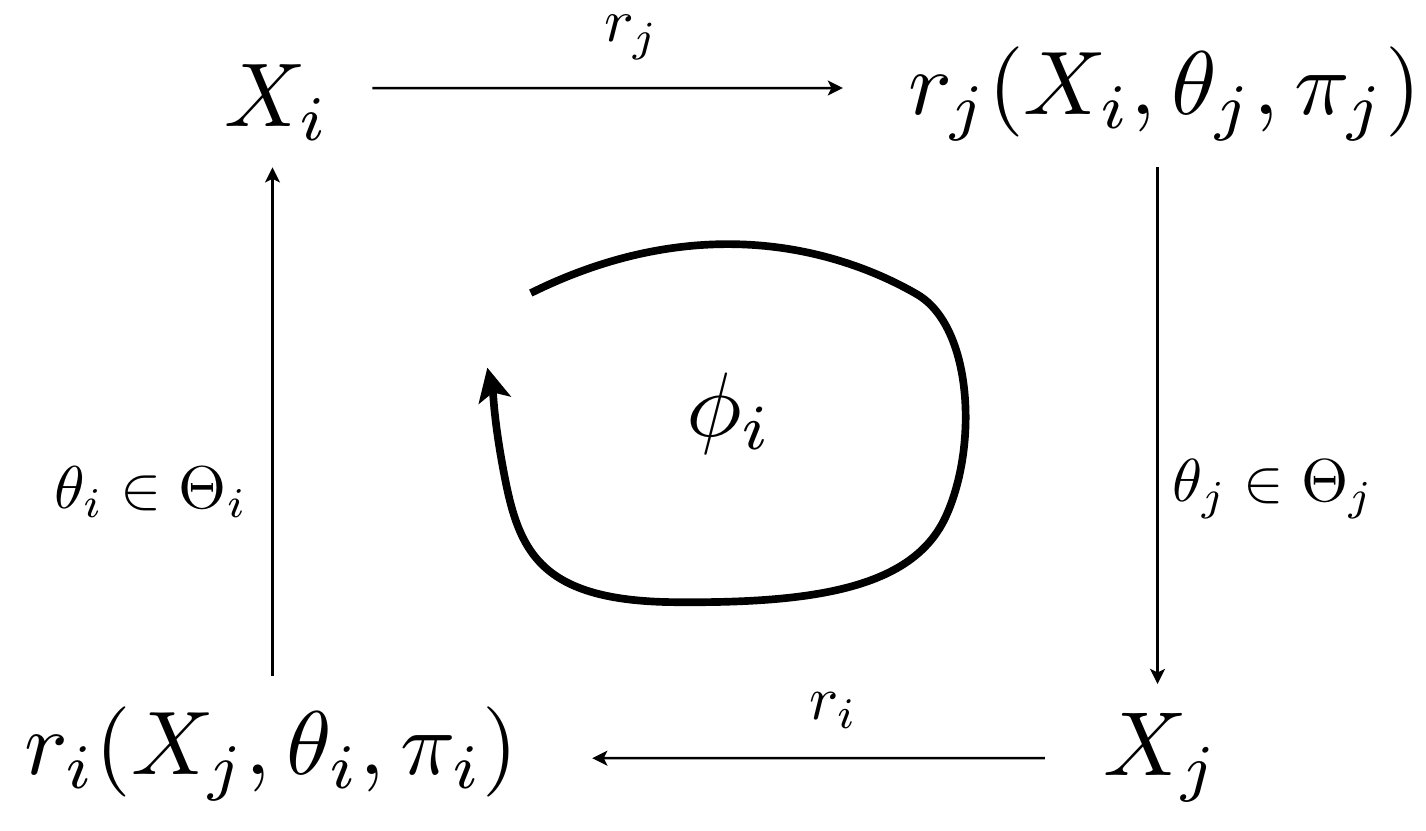}
	\caption{$\phi_i$ mapping}
	\label{fig:phi}
\end{figure}
\beno
	\underline{x}'_i = \arg \min_{\theta_i \in \Theta_i} r_i([\underline{x}_j, \overline{x}_j], \theta_i, \pi_i)\\
	\overline{x}'_i = \arg \max_{\theta_i \in \Theta_i} r_i([\underline{x}_j, \overline{x}_j], \theta_i, \pi_i)
\vspace{-0.1in}
\eeno
where
\vspace{-0.1in}
\beno
	\underline{x}_j = \arg \min_{\theta_j \in \Theta_j} r_j([\underline{x}_i, \overline{x}_i], \theta_j, \pi_j)\\
	\overline{x}_j = \arg \max_{\theta_j \in \Theta_j} r_j([\underline{x}_i, \overline{x}_i], \theta_j, \pi_j).
\eeno
From construction $\underline{x}'_i \le \overline{x}'_i$. If $\phi_i$ is a continuous mapping, then by Brouwer's fixed point theorem, there exists $(\underline{x}^*_i, \overline{x}^*_i) \in X_{i,o}^2$ such that
\beno
	\phi_i(\underline{x}^*_i, \overline{x}^*_i) = (\underline{x}^*_i, \overline{x}^*_i).
\eeno
Then $X_i = [\underline{x}^*_i, \overline{x}^*_i]$ is, by definition, an uncertainty equilibrium. Now we show that $\phi_i$ is continuous in $\underline{x}_i, \overline{x}_i$. 

Let $v:=y(\underline{x}_i, \overline{x}_i) :=\arg \sup_{x_i \in [\underline{x}_i, \overline{x}_i]} u_j(x_j, x_i, \theta_j)$ and define $z$ such that $\underline{x}_i-\epsilon \le z \le \underline{x}_i+\epsilon$. Then $\lim_{\epsilon \rightarrow 0} u_j(x_j,z,\theta_j) = u_j(x_j, \underline{x}_i, \theta_j)$ from $u_j$'s continuity. There are two cases: (1) $y(\underline{x}_i, \overline{x}_i) > \underline{x}_i$. Then $y(z, \overline{x}_i) = y(\underline{x}_i, \overline{x}_i)$ as for small $\epsilon$.  (2) $y(\underline{x}_i, \overline{x}_i) = \underline{x}_i$. Then $\underline{x}_i - \epsilon \le w:=y(z, \overline{x}_i) \le \underline{x}_i + \epsilon$. As a result
\beeno
	&\sup_{x_i \in [z, \overline{x}_i]} u_j(x_j, x_i, \theta_j) - \sup_{x_i \in [\underline{x}_i, \overline{x}_i]} u_j(x_j, x_i, \theta_j)&\\ &= u_j(x_j, w, \theta_j) - u_j(x_j, x_i, \theta_j) \rightarrow 0&
\eeeno
as  $\epsilon \rightarrow 0$ from $u_j$'s continuity. 

Therefore $\sup_{x_i \in [\underline{x}_i, \overline{x}_i]} u_j(x_j, x_i, \theta_j)$ is continuous in $\underline{x}_i$. 
Similarly we can show it is continuous in $\overline{x}_i$. These steps can be repeated for $\inf_{x_i \in [\underline{x}_i, \overline{x}_i]} u_j(x_j, x_i, \theta_j)$. 
As a result $f_j$ and $r_j$ are continuous in $\underline{x}_i, \overline{x}_i$. Since $r_j$ is continuous in $\theta_j$ and $\Theta_j$ is a closed and bounded interval, $X_j:=[\underline{x}_j, \overline{x}_j]:= \{r_j([\underline{x}_i, \overline{x}_i], \tilde{\theta}_j, \pi_j | \tilde{\theta}_j \in \Theta_j\}$ is a closed interval too. 
Using the same procedure, $\underline{x}'_i$ and $\overline{x}'_i$ are continuous in $\underline{x}_j, \overline{x}_j$. Since $\phi_i$ is a composite function of continuous functions in $\underline{x}_i, \overline{x}_i$, $\phi_i$ is therefore continuous in $(\underline{x}_i, \overline{x}_i)$. This completes the proof.

\subsection{Proof of Theorem \ref{thm:coincidence}}
Let $\Theta_i = \{ \theta_i \}$ for all $i$. Then for arbitrary $X_j$, $X_i:=\{r_i(X_j, \theta_i, \pi_i)| \theta_i \in \Theta_i\}$ is  a singleton. Let $X_i=\{x_i^\dagger\}$. Then 
\[
x_j^\dagger := r_j(X_i, \theta_j, \pi_j) =\arg \max_{x_j \in X_{j,o}} u_j(x_j, x_i^\dagger, \theta_j)
\]
is $j$'s best response function of game $\mathcal{G}_{o}$ when $j$ predicts $i$ plays $x_i^\dagger$. By assumption an equilibrium of this is a $(x_1^\star, x_2^\star)$. And by construction, it is also an uncertainty equilibrium $(X_1^*(\pi), X_2^*(\pi))$ of $\mathcal{G}(\pi)$, and it does not depend on $\pi$. 

\subsection{Proof of Theorem \ref{thm:no_pessimism}}
Suppose player 1's dominant attitude is pessimism. From Lemma \ref{lem:dominance}, this implies
\[
	\beta_{1} \ge \theta_{1} \ge \overline{\theta}_{1} = (2-\alpha_{1}+4\beta_{2}-2\alpha_{2})/3.
\]
Now then,
\beeno
	\overline{\theta}_{2} &=& (2-\alpha_{2}+4\beta_{1}-2\alpha_{1})/3 \\ &\ge& (14 - 10 \alpha_{1} - 11\alpha_{2} + 7\beta_{2})/9 + \beta_{2} > \beta_{2}.
\eeeno
Thus $\theta_{2} \le \beta_{2} < \overline{\theta}_{2}$. Therefore pessimism cannot be player 2's dominant strategy.

\subsection{Proof of Theorem \ref{thm:no_pessimism}}
Consider player 1 representatively. We will show $U_{1}(OO) > U_{1}(PO)$ for some $\theta_{2} \in \Theta_{2}$. 
Let  $u:=u_{i}$, $r:=r_{i}$ and $\Theta:=[\alpha, \beta] = \Theta_{i}$ for $i=1,2$. $\alpha < \beta$. As one case, assume $u_{i}$ is strictly decreasing in $x_{j}$, $r_{i}$ is decreasing in $x_{j}$ and $\theta_{i}$ both. The conclusion is the same if any of `decreasing' condition is changed to `increasing' condition.
Define equilibrium sets for each $\pi$ as follows:
\beeno
	&X_{1}=X_{2}=[a,b] \mbox{ for $\pi=(OO)$}\\
	&X_{1}=X_{2}=[c,d] \mbox{ for $\pi=(PP)$}\\
	&X_{1}=[e,f], X_{2}=[g,h] \mbox{ for $\pi=(OP)$}\\
	&X_{1}=[g,h], X_{2}=[e,f] \mbox{ for $\pi=(PO)$}.
\eeeno
Then
\beeno
	&a=r(a,\beta) \mbox{ and } b=r(a,\alpha)\\
	&c=r(d,\beta) \mbox{ and } d=r(d,\alpha)\\
	&e=r(g,\beta) \mbox{ and } f=r(g,\alpha)\\
	&g=r(f,\beta) \mbox{ and } h=r(f,\alpha).
\eeeno
From monotonicity of $r$, we draw relation one by one: From $a=r(a,\beta)$ and $d=r(d,\beta)$, it is immediate to see
$
	a < d.
$
Noting $d=r(r(d,\alpha),\alpha)$ and $g=r(r(g,\alpha),\beta)$, we get $g < d$. Thus 
$
	  d < f
$
from $d=r(d,\alpha)$ and $f=r(g,\alpha)$.
From $a < f$, we get $g < a$. Finally we get
$
	a  < e.
$
Take $\theta_{2}=\beta$. Then,
\beeno
	U_{1}(OO) &=& u(x_{1}(\theta_{1},OO), x_{2}(\theta_{2}, OO), \theta_{1})\\
	&=& u(r(a,\theta_{1}), r(a,\theta_{2}), \theta_{1})\\
	&=& u(r(a,\theta_{1}), r(a,\beta), \theta_{1})\\
	&=& u(r(a,\theta_{1}), a, \theta_{1})\\
	&>& u(r(f,\theta_{1}), a, \theta_{1})\\
	&>& u(r(f,\theta_{1}), e, \theta_{1}) = U_{1}(PO)
\eeeno
Therefore pessimism cannot be a dominant attitude in a symmetric game.

%


\bibliographystyle{abbrv}
\bibliography{attitude}

\end{document}